# Engineered optical nonlinearities and enhanced light transmission in soft-matter systems with tunable polarizabilities

Weining Man[1*], Shima Fardad [1,2*], Ze Zhang[1-3*], Jai Prakash[1], Michael Lau[1], Peng Zhang[1,4], Matthias Heinrich[2], Demetrios N. Christodoulides[2], and Zhigang Chen[1,5]

[1]Department of Physics and Astronomy, San Francisco State University, San Francisco, CA 94132
[2]CREOL/College of Optics, University of Central Florida, Orlando, FL 32816
[3]Academy of Opto-electronics, Chinese Academy of Science, Beijing 100094, China
[4]NSF Nanoscale Science and Engineering Center, University of California, Berkeley, CA 94720
[5]TEDA Applied Physics School, Nankai University, Tianjin 300457, China
* These authors made equal contribution.
zhigang@sfsu.edu, demetri@creol.ucf.edu

**Synthesizing artificial material systems exhibiting unique optical properties not found in nature is nowadays one of the major scientific endeavors[1-5]. This challenge becomes particularly acute in soft-matter environments where light scattering loss has always been a formidable hurdle[6]. In the domain of nonlinear optics, this issue becomes further complicated in colloidal suspensions where typically the corresponding optical polarizability is positive[7]. Under such conditions, enhanced scattering takes place as optical beams tend to catastrophically collapse[8] – causing a dramatic drop in transmission. Clearly, of importance will be to devise approaches to overcome these limitations. If successful, one could envision stable low-loss propagation of nonlinear needles of light, which might in turn be used to non-invasively initiate and control chemical and mesoscopic kinetic processes[9-11]. In this work, we demonstrate that the nonlinear response of certain soft-matter systems can be tailored at will by appropriately engineering their optical polarizability. In particular, we deliberately synthesize stable colloidal suspensions with *negative* polarizabilities, and observe for the first time robust propagation and enhanced transmission of self-trapped light over long distances that would have been otherwise impossible in conventional suspensions with positive polarizabilities. What greatly facilitates this behavior is an induced saturable nonlinear optical response introduced by the thermodynamic properties of these colloidal systems. This in turn leads to a substantial reduction in scattering via self-activated transparency effects. Our results may open up new opportunities in developing soft-matter systems with tunable optical nonlinearities.**

Controlling light transport in soft-matter systems could be crucial in many and diverse fields of science and technology. For example, in colloidal suspensions, this can be accomplished through optical radiation forces capable of manipulating particle concentration and molecular kinetics at the mesoscopic level[2,7,11,12]. In principle, such optically induced processes can be exploited for initiating and regulating chemical reactions, for sorting different species of nanoparticles, and for influencing diffusion and osmotic pressure effects, to mention a few[13]. Capabilities of this sort may have important ramifications in life sciences, where for instance handling and separation of cells or viruses is often necessary[14]. So far however, meeting these goals has been challenging. Not surprisingly, what has impeded progress in this area has much to do with the underlying physics of light propagation in optically non-uniform media. In random environments, light scattering dominates, thus greatly diminishing transmission. Given that the polarizability of most stable colloids happens to be positive, particles tend to migrate into the high-intensity regions of an optical beam—making propagation conditions even worse because of enhanced scattering[2]. At the same time, this situation is further exacerbated by optical catastrophic self-focusing collapse initiated by a non-saturable Kerr nonlinearity–originating from such positive polarizability arrangements[15]. In view of this, the question naturally arises whether one could overcome these obstacles by judiciously crafting the nonlinear response of a colloidal system. If so, what are the key physical variables in controlling this behavior?

These issues can be addressed by first considering how the particle polarizability is related to optical nonlinearities that are solely mediated by radiation pressure effects. In general, a particle displays a positive polarizability (PP) whenever its refractive index exceeds that of the background medium, while in the converse case its polarizability is negative (NP). As indicated in several studies[16], under the action of gradient forces, PP dielectric particles are attracted towards the center of an optical beam where the intensity is higher, whereas their NP counterparts are repelled [as illustrated in Fig. 1(a) and 1(b)]. In view of these dynamics, one can readily conclude that the refractive index of a colloid will in both cases increase along the path of the beam– always resulting in a self-focusing nonlinearity[17]. It is worth noting that until recently, the common belief has been that PP arrangements exhibit a pure Kerr nonlinearity, in which case the refractive index change is expected to vary linearly with intensity. Yet, recent z-scan measurements have revealed that the PP nonlinearity exceeds the typical Kerr response because of the exponential character of the Boltzmann law involved in the particle distribution function[18]. This super-critical nature of the nonlinearity explains why in typical experimental settings optical beams quickly become highly unstable and catastrophically collapse in PP suspensions[15]. A possible way to overcome this problem was proposed in a recent theoretical study where it was indicated that NP suspensions should instead exhibit a *saturable* Kerr response[17]. If indeed this is so the case, then one could expect stable optical self-trapping and self-induced transparency, entirely free of the aforementioned complications. In other words, by appropriately engineering the polarizability of such arrangements one can tailor their nonlinearity. Furthermore, by deliberately mixing PP and NP suspensions [as illustrated in Fig. 1(c)], one could synthesize soft-matter systems with a tunable polarizability, optimized nonlinear response, and enhanced light transmission.

In this study, we experimentally demonstrate a new class of synthetic colloidal suspensions capable of exhibiting negative polarizabilities. This is accomplished in a stabilized mixture of Polytetrafluoroethylene (PTFE) particles in a glycerin-water solution. We show that by judiciously introducing NP particles in conventional PP colloidal suspensions, the resulting "mixed" polarizability can be fine-tuned, thereby enabling us to modify the nonlinear response of these systems. In particular, we observe robust propagation and up to a fourfold-enhanced transmission of an optical beam when traversing an NP suspension as compared to that in a typical PP suspension. Such light penetration through otherwise strong scattering environment is attributed to the interplay between optical forces and self-activated transparency effects while no thermal effect is involved. Our experimental observations are in agreement with a previously derived thermodynamic model that takes into account the interplay between the optical intensity and the osmotic pressure in the presence of particle-particle interactions[19]. These findings may pave the way towards synthesizing soft-matter systems with customized optical nonlinearities that can enable an intensity-dependent reduction in scattering losses.

To understand light-particle dynamics in colloidal dispersions, we first consider the optical gradient force that typically dominates the interaction process. To first order, this component of the optical force is given[2,16] by $\vec{F} = \alpha \nabla I/4,$ where $I$ is the optical intensity and $\alpha$ is the polarizability of the particle. In the dipole regime, $\alpha = 3V\varepsilon_0 n_b^2 (m^2 - 1)/(m^2 + 2)$, where $V$ is the particle's volume, $m = n_p/n_b$ is a measure of the contrast between the refractive index of the particle $n_p$ and that of the background fluid $n_b$. Therefore, PP dielectric particles with $n_p > n_b$ ($\alpha > 0$) will be attracted towards the high intensity regions of an optical beam (Fig. 1(a)), whereas NP particles with $\alpha < 0$ ($n_p < n_b$) will be repelled (Fig. 1(b)). A mixed response is expected when both species are present (Fig. 1(c)). To visualize these phenomena experimentally, we launch a focused CW laser beam ($\lambda = 532$ nm) into three samples of aqueous colloidal suspensions ($n_b = 1.33$), which are comprised of PP, NP, and a mixture of PP and NP particles, respectively. To allow for a direct observation of these effects we here use macroscopic particles. Specifically, for the PP case we employ 2 μm polystyrene beads ($n_p = 1.59$) suspended in water while in this same environment, hollow silica spheres (with an average size of 7 μm) serve as NP particles. Typical experimental results are depicted in Figs. 1(d-f), and corresponding real-time movies showing the resulting dynamics are provided in the Supplementary Information S1. In the PP suspension, optical gradient forces overcome Brownian motion and hence the particles are strongly attracted towards the beam center (Fig. 1(d)). Conversely, in the NP regime, the low-index particles are expelled from the beam path as clearly indicated in Fig. 1(e). Finally, when both types are present, the particles individually respond to this optical stimulus and are therefore spatially separated within the beam (Fig. 1(f)). We note that even though the hollow silica spheres can provide an NP setting, they cannot be readily scaled down in the optical subwavelength regime—a necessary attribute for low scattering suspensions.

In what follows we study the optical properties of a specially synthesized NP nanosuspension. To this end, we suspend polytetrafluoroethylene (PTFE) particles (200 nm in diameter) with $n_p = 1.35$ in a glycerin-water mixture (at 3:1 ratio) having an average index of $n_b = 1.44$ and a viscosity of 55 mPa·s. This colloid was obtained by diluting a commercially available high-concentration PTFE aqueous solution and is sterically stabilized to avert coagulation (see Methods). In a first set of experiments, a laser beam ($\lambda = 532$ nm) is launched into 10-mm-long glass cuvettes filled with these suspensions. Both the linear and nonlinear propagation dynamics are observed by monitoring scattered light using a side-view camera. A schematic illustration and description of our experimental setup is given in the Supplementary Information S2. For a fair comparison, an equivalent PP colloidal suspension based on functionalized polystyrene (PS) particles ($n_p = 1.59$) of the same average size (200 nm) was realized in the same liquid background.

In order to establish a proper reference for our studies, we first perform experiments in the PP polystyrene colloidal suspension. In all cases the laser beam enters the sample at its minimum waist, having an intensity full width at half maximum (FWHM) of 11 μm. At low power levels this beam considerably expands because of linear diffraction, after propagating over 14 diffraction lengths. Yet, above a certain power threshold, nonlinear self-focusing effects overcome this broadening. Specifically, at a power of 3 W, the beam undergoes catastrophic self-focusing collapse after only a few millimeters of propagation (Fig 2 (a)). This situation becomes further complicated by severe scattering losses accompanying this same process. As a result, the optical beam quickly destabilizes and dissipates its energy. As outlined above, what leads to this response is the super-critical nature of the self-focusing nonlinearity associated with a PP suspension, as well as the fact that in this same regime the concentration of the scattering particles tends to increase with intensity. To better understand this behavior we have to first examine how this optical nonlinearity arises from mesoscopic kinematics. To do so, one has to allow for particle-particle interactions through a virial expansion in the osmotic pressure of this "non-ideal gas". The virial coefficients involved heavily depend on the specific stabilization mechanism (electrostatic or entropic) as dictated by van der Waals forces and the corresponding Debye-Hückel model[6]. By introducing the corresponding compressibility factor in the generalized Fick's law and by assuming equilibrium conditions, the optical intensity $I$ and the perturbed volume filling factor $f$ are then found to be related as follows[19]:

$$\frac{\alpha I}{4k_B T} \simeq \ln\frac{f}{f_0} + \frac{2B_2 f_0}{V}\left(\frac{f}{f_0} - 1\right) + \frac{3B_3 f_0^2}{2V^2}\left(\left(\frac{f}{f_0}\right)^2 - 1\right) \ . \qquad (1)$$

Equation (1) describes the nonlinear response of the colloidal system. Here, two- and three-body interactions have been accounted for, as reflected by the presence of the virial coefficients $B_{2,3}$. A closer look at Eq. (1) reveals that, in the PP case ($\alpha > 0$), a super-critical response emerges as result of the logarithmic Boltzmann term and the two-particle interaction ($B_2$). This explains the observed rapid beam collapse shown in Fig. 2(a). If on the other hand, the polarizability is negative ($\alpha < 0$), then after a characteristic time, the particles are expelled from the beam, hence locally diminishing the role of the terms associated with $B_{2,3}$. Consequently, the NP behavior is now dominated by the Boltzmann law $f = f_0 \exp(-|\alpha|I/4k_B T)$. This latter expression predicts that the NP self-focusing nonlinearity should instead be saturable, in stark contrast to the PP case. Note that this saturable nature of the nonlinearity is essential to avoid beam collapse during propagation[17-20]. In other words, in an NP environment one could anticipate self-trapped "needles of light" that are capable of propagating over long distances. As we will see, the longevity of these nonlinear entities can be further enhanced in NP systems because of reduced scattering losses (Supplementary Information S3).

To verify this hypothesis we launch the same optical beam (3 W, 11 μm) in the polytetrafluoroethylene NP suspension. Although the particle size, the initial volume filling factor, and the liquid background remain the same, Figure 2(b) indicates clearly that, in this NP regime, the optical beam can more effectively penetrate this scattering soft-matter arrangement in a stable manner. In contrast to the PP case where the injected Gaussian wave front disintegrates after few millimeters, in the NP setting it maintains its cross section and experiences considerably lower losses. To demonstrate that the observed self-trapping behavior is indeed mediated purely by optical forces and not caused by any thermal effects, we have also prepared an index-matched suspension where the PTFE particles were suspended in a dilute glycerin-water mixture with $n_b = n_p = 1.35$. Under these conditions, the polarizibility $\alpha = 0$, and thus no gradient forces can act on the particles. Our experiment shows that up to power levels of 3 W, the beam broadens by a factor of 14, as expected from linear diffraction theory (Fig. 2(c)). Given that both water and glycerin exhibit a substantial negative temperature coefficient in their refractive index ($dn/dT \sim 10^{-4} K^{-1}$) and that no self-defocusing is detected in the range of powers investigated, one is compelled to conclude that no thermal effects are at play in our system. To examine the stability of the nonlinearly induced self-trapped beams, we additionally monitor their width upon propagation. To this end, we use shorter cuvettes (2 mm) to minimize the scattering background during imaging. Figure 3(a) shows the beam profile and its corresponding cross-section at the input. At lower powers (linear regime), the beam expands from 11 μm to 32 μm, which is consistent with ~3 diffraction lengths (Fig. 3(b)). At a higher power (3W) the beam is found to maintain its initial width (Fig. 3(c)). This conclusively illustrates stable soliton formation[20] in this NP system. Figure 3(d) displays the propagation of such a self-trapped beam over a distance of 5 mm. We emphasize again that stable self-trapping supported by NP nonlinearities is fundamentally different from that observed in PP arrangements[21,22], such as those due to electrostrictive nonlocal response [23,24], optically induced thermodiffusion, and thermophoresis effects[25,26].

The successful demonstration of NP behavior now allows one to synthesize soft-matter systems with tailored optical nonlinearities. This can be done by judiciously mixing both PP and NP particles within the same dispersant. In such a setting, the resulting self-focusing arises from the collective action of both species involved—with PP particles being attracted and NP ones being repelled by the beam. Yet, despite the fact that both processes lead to an increase in the refractive index, the respective Boltzmann terms (in Eq. (1)) contribute to the nonlinearity in entirely different ways. As a result, a gradual inclusion of NP particles into an initially pure PP system can transform its super-critical response into a saturable one. This scenario is depicted in Fig. 4(a) along with the "Kerr plane" marking the transition between these two regimes. Experiments confirming these prospects were carried out in mixtures of PTFE and PS particles. As Fig. 4(b) shows, the beam FWHM tends to decrease as more PP particles are introduced into an NP suspension (0.7%)— indicating that while the net nonlinearity increases, self-focusing remains stable. Yet, once the polystyrene content exceeds 0.3%, the system enters the supercritical phase and hence the beam undergoes catastrophic collapse. We next investigate self-induced transparency effects in NP dispersions. In general, scattering losses in colloidal systems increase with concentration. Given that in NP arrangements, the filling factor decreases exponentially with intensity, the overall transmission nonlinearly increases. In other words, in an NP setting, an intense beam should be capable of clearing up the haze from its own path—a desirable feature in scattering environments. Along these lines, we experimentally observe a substantial enhancement in transmission (from 18% to 34%) when the power was increased to 1.6 W in a pure 0.7% PTFE NP dispersion (Fig. 4(c)). This is in stark contrast to what we measure in an equivalent polystyrene PP suspension having identical linear losses. Here the transmission drops by a factor of two (from 18% to 10%) at the same power level. An intermediate response is observed in mixed suspensions.

In summary, we have proposed and demonstrated that the nonlinear response of soft-matter systems can be engineered at will by appropriately tailoring their optical polarizability. This was accomplished by synthesizing colloidal suspensions with negative polarizabilities. Our results may open up new opportunities in developing versatile soft-matter systems with tunable optical nonlinearities. We have shown that by introducing NP particles into a PP environment one can alter the colloidal nonlinearity—thus allowing stable self-trapped "needles of light" over substantial distances. Interestingly, this is further enabled by a nonlinearly activated self-induced transparency effect. We envisage that, by locally modifying the particle concentration, such light filaments could be instrumental in controlling mesoscopic kinetic processes like osmosis, diffusion, etc. along their path. Particles coated with catalysts, enzymes, or reactants, optically manipulated to either promote or inhibit specific chemical reactions, could be another example. In addition, our work might bring about new possibilities for the dynamically changing fields such as optofluidics[9] and disordered photonics[27], where understanding optical manipulation and transmission of light through scattering media is certainly desirable.

**Acknowledgements**

This work was supported by the Air Force Office of Scientific Research (MURI, grants FA9550-10-1-0561, FA9550-12-1-0148, FA9550-12-1-0111) and by NSF (grants ECCS-1128520, PHY-1100842). MH was supported by the German National Academy of Sciences Leopoldina (grant LPDS 2012-01). The Authors would like to thank Dr. Alessandro Salandrino for valuable discussions.

**Figures:**

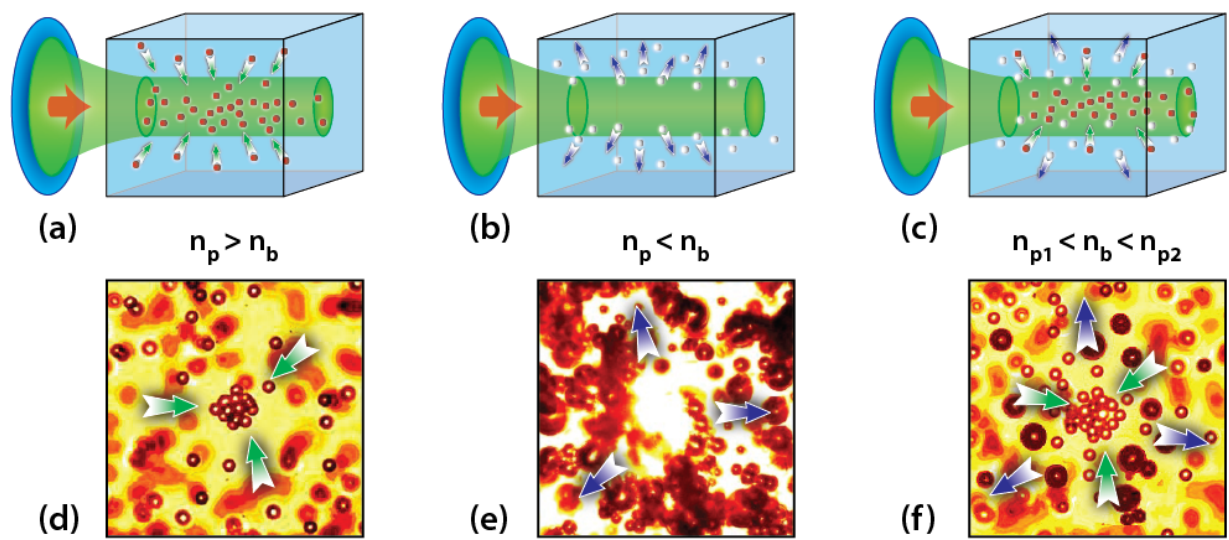

**Figure 1:** Schematic illustrations (a-c) and experimental demonstrations (d-f) of light-particle interactions in colloidal suspensions. (a,d) Positive-polarizability particles (2 μm polystyrene beads suspended in water) are attracted by an optical beam. (b,e) Repulsion of negative-polarizability particles (7 μm hollow glass spheres suspended in water). (c,f) Collective motion of PP and NP particles when both species are present in a colloidal mixture. Movies of the three cases can be found in the Supplementary Information S1.

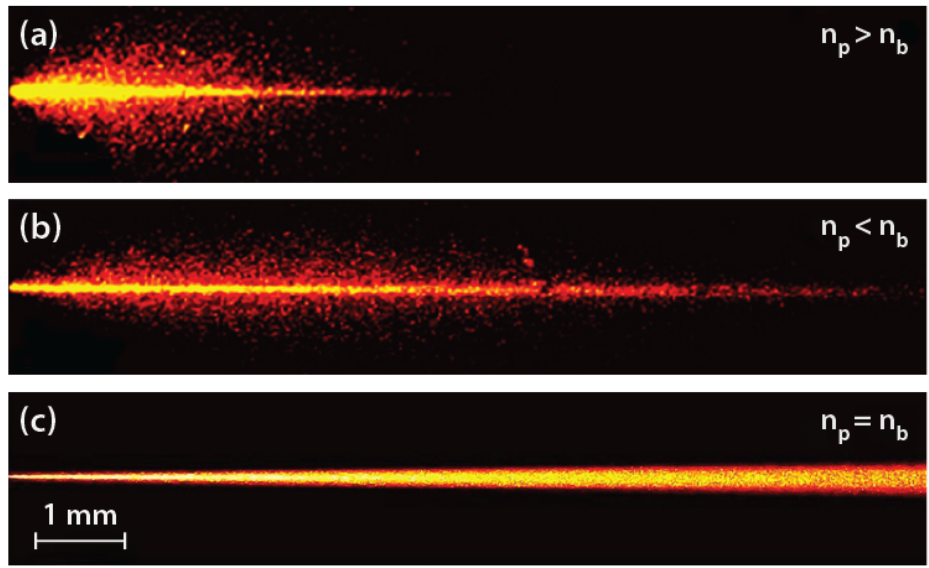

**Figure 2**: Experimental observation of a light beam propagating through 10-mm-long colloidal suspensions of different polarizabilities. (a) Catastrophic self-focusing collapse occurs in the PP suspension. (b) Enhanced transmission against scattering is observed in the NP suspension over several diffraction lengths. (c) The same beam diffracts linearly even at high power in a refractive index matched (zero-polarizability) suspension. In (c) the image contrast has been enhanced to improve visibility despite the much weaker scattering.

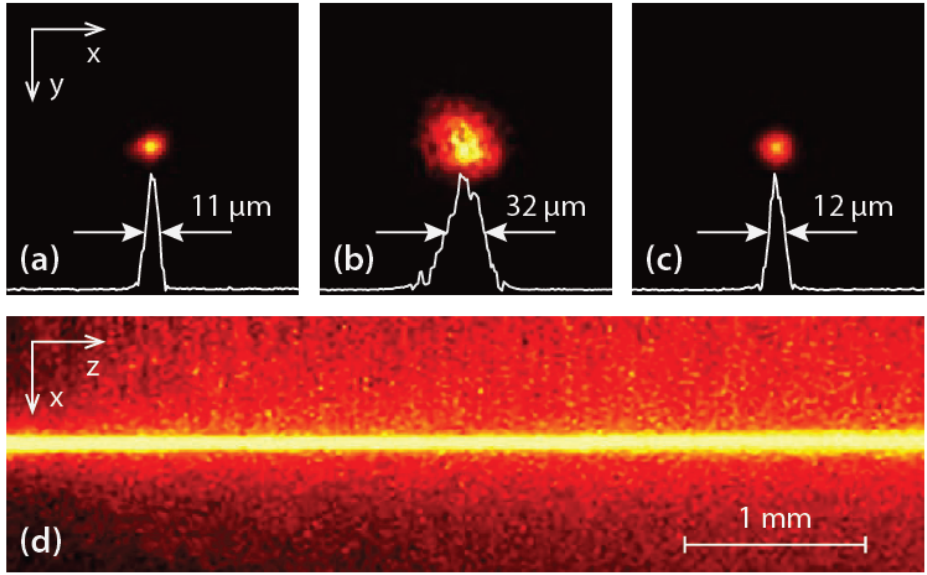

**Figure 3**: Observation of stable optical self-trapping in an NP colloidal suspension with $f_0 = 0.3\%$. (a) Input beam profile; (b) Linear (low power) output pattern after a propagation distance of 2 mm; (c) Nonlinear output intensity profile corresponding to a stable soliton at 3 W. The side-view photograph (d) shows the propagation of such a self-trapped beam over a distance of 5 mm.

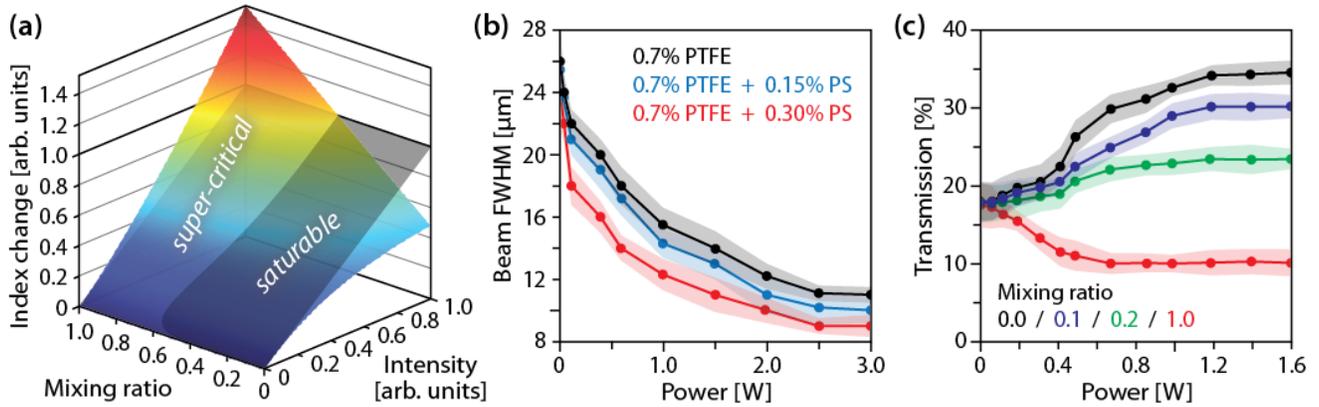

**Figure 4**: (a) Normalized nonlinear response of a mixed-polarizability suspension. The collective nonlinearity arises from a superposition of the two components for a fixed overall volume filling factor. The pure NP dispersion corresponds to a mixing ratio of 0, while a ratio of 1 indicates a pure PP system. Despite the fact that both species provide a positive nonlinear index change, the saturable contribution from the NP component can serve to balance the supercritical PP behavior. The transition between the two regimes is indicated by the Kerr plane (gray). (b) Measured output beam size after 2 mm of propagation as a function of the input power for mixed colloidal suspensions containing 0.7% of NP and varying amounts of PP particles. Self-trapping is enhanced by the presence of the PP component but remains stable only up to 0.3% of PP particles, above which catastrophic beam collapse occurs. (c) Measured intensity-dependent transmission for different suspensions (black: pure NP; red: pure PP; blue, green: mixed) after 2 mm of propagation. All mixtures were appropriately diluted to exhibit the same linear scattering loss of 18%. Self-induced transparency effects are clearly evident in these systems. In (b) and (c) the error ranges are indicated by the shaded regions surrounding the respective graphs.

**Methods**

*Colloidal suspensions*

To synthesize colloidal suspensions with a negative polarizability, we dispersed polytetrafluoroethylene particles (diameter $d = 200$ nm, refractive index $n_p = 1.35$) with various volume filling factors ranging between 0.3 and 0.7% in a 3:1 glycerin-water solution having a background refractive index of $n_b = 1.44$. Positive-polarizability suspensions were prepared from polystyrene particles ($n_p = 1.59$) of the same size. The NP suspensions used in this study were prepared by diluting a commercially available high-concentration PTFE colloidal suspension (Ultraflon A-100, Laurel Products).

*Time scales involved*

The 3:1 glycerin-water mixture used as dispersant had a viscosity[28] of $\eta = 5.5 \cdot 10^{-2}$ Pa·s. For particles with a mass of $M \approx 10^{-27}$ kg, as used in our experiments, the resulting dynamic Langevin time scale[13] of the suspension is $\tau = M/3\pi\eta d \approx 10^{-10}$ s. However, the optical response is governed by the driven particle transport into and out of the beam (having a spatial extent $w$). In this case, the corresponding equilibrium time scale of the colloidal system is given by $\tau_{eq} = 3\pi n_b w^4 \eta d / 2Z_0 P_0 \alpha$, where $Z_0 = 377\ \Omega$ is the free space impedance. Taking into account that in our case the beam has a width of $w = 11$ μm, and carries an optical power $P_0$ of up to 3W, while the particles exhibit a polarizability of $|\alpha| \approx 10^{-32}$ F·m², we estimate this response time to be on the order of seconds. Note that, as $\alpha$ is proportional to the volume $V$, $\tau_{eq}$ scales with the inverse square of its diameter $d$ of a particle, thereby readily allowing for an adjustment of the suspension's optical response time if so desired.